## MOLECULAR SPECTROSCOPY

# Simulation of Time-Resolved Vibronic Spectra and the Possibility of Analyzing Molecules with Similar Spectral Properties

S. A. Astakhov and V. I. Baranov

*Institute of Geochemistry and Analytical Chemistry, Russian Academy of Sciences, Moscow, 117975 Russia*
*e-mail: baranov@geokhi.ru*
Received March 23, 2001

**Abstract**—The possibility of using time-resolved vibronic spectroscopy for spectral analysis of mixtures of chemical compounds with similar optical properties, when traditional methods (based on stedy-state spectra) are inefficient, is demonstrated by using the method of computer simulation. The analysis is carried out by the example of molecules of a series of polyenes (butadiene, hexadiene, octatetraene, decapentaene, and decatetraene), their various *cis*- and *trans*-rotational isomers, and phenyl-substituted polyenes. Ranges of relative concentrations of molecules similar in their spectral properties, where reliable interpretation of time-resolved spectra of mixtures and both qualitative and quantitative analyses are possible, are determined. The use of computer simulation methods for optimizing full-scale experiments in femtosecond spectroscopy is shown to hold much promise. © 2002 MAIK "Nauka/Interperiodica".

## INTRODUCTION

The use of experimental facilities for recording fine-structure absorption and fluorescence spectra with high (picosecond and femtosecond) temporal resolution is one of the promising lines of development of modern vibronic spectroscopy [1, 2]. Interpretation of new experimental results with the aim of extracting information on molecular properties requires the construction of corresponding theoretical models of molecules, the development of methods for direct calculation of time-resolved spectra, and the execution of numerical experiments.

The possibility of simulating the time-resolved vibronic spectra of polyatomic molecules was analyzed in [3]. A method for calculating such spectra using molecular models described by parameters, which are not directly related to any particular spectral experiment (potential surfaces of ground and excited states in natural coordinates), was suggested [3]. In addition, software for personal computers allowing fast (real time) calculations of time-resolved spectra to be carried out for complicated molecules containing several tens of atoms as well as for their mixtures and including the construction of complete three-dimensional spectral curves as well as for their various sections on the frequency and time scales was developed [3]. An approved semi-empirical method ensuring a quantitative prediction of vibronic spectra with a high accuracy (frequency and intensity errors for the most typical fundamental lines are $\Delta\omega \leq 20–30$ cm$^{-1}$ and $\Delta I \leq 15–20\%$, respectively) is used for calculating the parameters of adiabatic potential surfaces in excited states [4–8].

It is known that the structure of ordinary (without time resolution) absorption and fluorescence spectra of a number of molecules (for instance, polyenes, phenyl- and diphenylpolyenes), their stereoisomers, and substituted forms (for instance, *cis*-, *trans*-isomers, deutero-, methyl-, and phenyl-substituted) often have a very similar form (see, for instance, data in [4, 5, 7, 9–15]). Differences in the intensities of the main vibronic lines (~10%) are comparable with the accuracy of their calculation. Therefore, interpretation and analysis of such spectra and, more importantly, identification of compounds (particularly in the case of mixtures) on the basis of spectral data is extremely complicated, which significantly restricts the possibility of using spectral methods for solving problems of analytical chemistry and photochemistry (in particular, photoisomerization processes [16]).

It was shown that the use of time-resolved vibronic spectra makes it possible to pose and to solve problems of qualitative and quantitative analyses of complex compounds with similar spectral properties without using standards [17, 18]. For instance, for comparable intensities of the main lines of spectra ($0.1 < I'_i/I''_j < 10$) of a two-component mixture, concentrations can be determined with confidence for a wide range of values (on the order of 1–110%) if their time dependences differ by a factor of 3 [18].

It is interesting to study the possibility of using time-resolved vibronic spectroscopy in molecular modeling and the solution of spectrochemical problems for various isomeric and substituted forms of complex compounds by methods of computer simulation (numerical experiment). The execution of these numerical experiments is very important because, on the one hand, it allows us to develop computational methods and to determine their potentials more clearly. On the other hand, the execution of a full-scale spectral experiment is rather laborious and requires preliminary analysis and simulation.

Molecules of a series of polyenes and their phenyl-substituted forms were used as the objects of this analysis because their spectral properties and the structure of excited states are of interest in their own right, in particular, in photochemistry (see, for instance, [9, 10, 16]) and are well reproduced by developed theoretical methods [5, 6, 8, 14, 15].

## METHOD FOR CALCULATING THE SPECTRA

For simulating the time-dependent vibronic spectra of polyenes, their phenyl- and methyl-substituted forms, and various *cis*- and *trans*-isomers, we used the method suggested and described in detail in [3]. The principle of the method is as follows.

A molecule is assumed to be excited by a short laser-radiation pulse much shorter in duration than the characteristic relaxation time of excited states. Once an absorption event comes to an end (the start of time counting), the time dependence of the intensity of each spectral line at the frequency $\omega_{ij}$ is proportional to the probability $w_{ij}$ of the corresponding vibronic transition from the state $i$ to the state $j$ and to the population of the initial level $n_i$

$$I_{ij}(\omega_{ij}, t) = \hbar\omega_{ij}w_{ij}n_i(t).$$

Time-resolved spectra $I_{ij}(\omega_{ij}, t)$ were calculated ignoring nonradiative transitions; therefore, the quantities $w_{ij}$ representing probabilities of optical vibronic transitions in this case were determined by the parametric method with the use of molecular models and fragments from the special data bank [4, 5]. The calculation was carried out in the Franck–Condon approximation for all vibronic transitions between lower states up to the fifth electron-excited state. Vibrational quantum numbers took values up to $v_{max} = 3$ because the result did not change significantly for $v_{max} > 3$. Not only overtones were taken into account, but composite vibrations as well.

Time dependences of the populations $n_i(t)$ of vibronic levels were found as the solutions to the corresponding system of rate equations

$$\frac{dn_i}{dt} = -w_i n_i + \sum_{j=i+1}^{N} w_{ji}n_j, \quad i = 1, 2, ..., N, \quad (1)$$

where $w_i = \sum_{j=1}^{i-1} w_{ij} + w_{i0}$ are total probabilities of transitions from the $i$th state to all lower lying excited states and the ground ($w_{i0}$) state.

By virtue of a high dimensionality of system (1) (up to $N \sim 30000$ in this case) and a wide range of values of $w_i$ (in particular, degenerate, $w_i = w_k$), the numerical method developed in [3] was used for its solution.

To make the analysis illustrative and to compare the spectral data obtained, we presented the simulation results for two-component mixtures of molecules under study and in the form of three-dimensional spectral surfaces and their various sections in the time and the frequency axes.

## RESULTS OF NUMERICAL EXPERIMENTS

Simulation was carried out for butadiene, hexadiene, octatetraene (OT), decapentaene, and decatetraene molecules, their various *cis*- and *trans*-rotational isomers, and phenyl-substituted derivatives. Consider the results obtained by using a number of typical examples.

Ordinary (stationary) fluorescence spectra of rotational isomers of polyenes, as noted above, differ little, mainly in the intensity of the line ~1200 cm$^{-1}$ corresponding to the vibration of single C–C bonds. This is confirmed, in particular, by the results of *ab initio* calculations [19]. For instance, such spectra for two rotational isomers of octatetraene are shown in Fig. 1a. Although identification of each isomeric form from the spectrum is basically possible (for instance, progression in the vibration of frequency 1200 cm$^{-1}$ for *trans*-octatetraene is approximately twice as intense as for *cis*-octatetraene; lines of the first components of this progression are marked by asterisks in Fig. 1a), this identification can be strongly hampered by such factors as significant broadening of spectrum lines and their superposition, low spectral resolution, and the presence of impurities (other isomeric forms can act as such impurities), [10, 11] all the more so since the total intensity of the components ~1200 and ~1600 cm$^{-1}$ varies by much less, ~10% (here and below, the intensity of the 0–0 transition is taken as 100%). Since energies of pure electron transitions of these isomers are close to each other (34300 and 36200 cm$^{-1}$), spectra of both isomeric forms in the mixture will be superimposed in the case of their simultaneous emission, and the assignment of lines of stationary experimental spectrum to each isomer turns out to be a difficult problem even for fine-structure spectra.

Time-resolved fluorescence spectra of these molecules turn out to be fundamentally different because the decay times of spectral lines of *cis*- and *trans*-isomers differ significantly (for instance, 130 and 50 ps for *cis*- and *trans*-octatetraene, respectively). Three-dimensional spectrum of a mixture of *cis*- and *trans*-octatetraene and its sections at various instants of time is presented in Fig. 2 as an example. It is clearly seen that the spectrum structure significantly changes with time and the lines belonging to the spectra of different isomeric forms of molecules are easily identified (the most intense lines of those belonging to the *trans*-form are marked by asterisks in Fig. 2). As shown in [18], this great difference in time dependences allows the spectra of various components of the mixture to be reliably separated and qualitative and quantitative analyses to be carried out in a very wide range of relative concentrations (from $C_{cis\text{-}OT}/C_{trans\text{-}OT} = 0.45$ to $C_{cis\text{-}OT}/C_{trans\text{-}OT} = 45$ in this case).

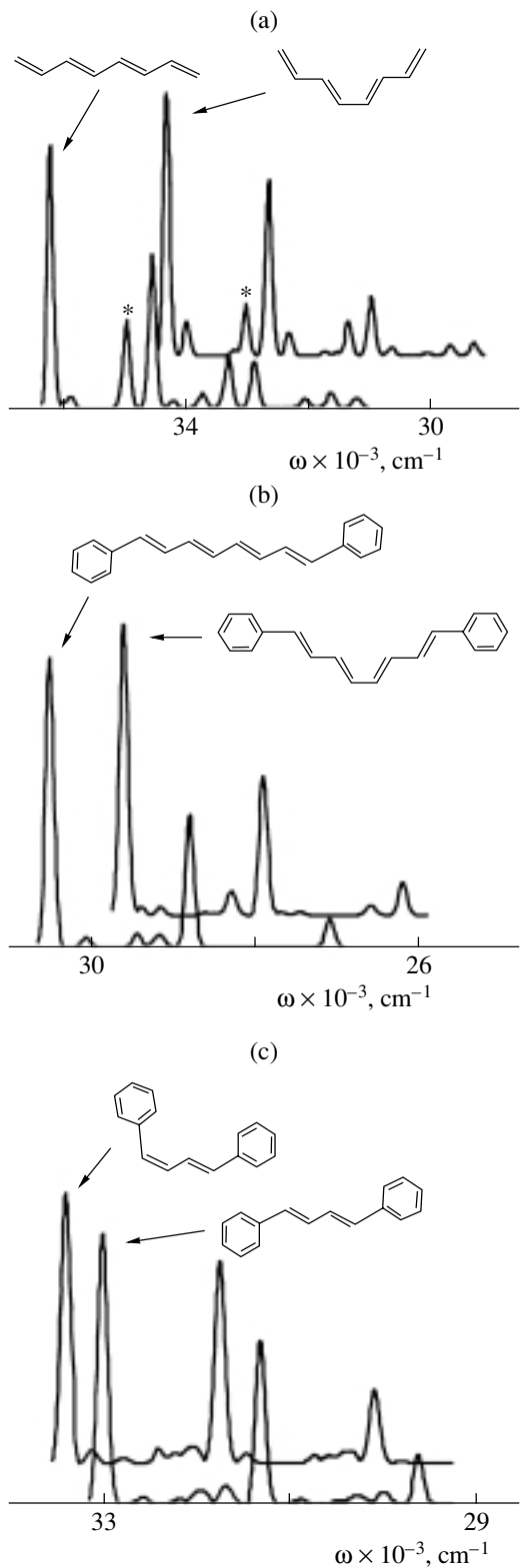

**Fig. 1.** Stationary fluorescence spectra ($S_1 \longrightarrow S_0$) of models of rotational isomers of (a) octatetraene, (b) diphenyloctatetraene, and (c) phenyl-substituted butadiene.

Henceforth, no greater than a tenfold difference in the intensities of the first and the second components of the mixture spectrum at the initial instant of time ($0.1 < I_1(0)/I_2(0) < 10$) and the requirement that their ratio should change by no less than a factor of $e$ during the time of spectrum observation comparable with the lifetimes of electron-excited states are taken as necessary conditions for this interpretation for estimating the concentration ranges. The first requirement corresponds to the appearance of vibronic lines of both components in the mixture spectrum observed. The second requirement corresponds to a significant variation of their relative intensities needed for interpretation (see [18] for more detail).

Energies of pure electron transitions in isomers differ insignificantly as a rule, by one or two vibrational quanta (~2000 cm$^{-1}$ for *cis*- and *trans*-octatetraene). However, in the case of simultaneous excitation of molecules in the mixture by a pulse with a small spectral width, different vibration sublevels of isomers become populated. For instance, the spectrum in Fig. 2 corresponds to initial excitation of a pure electron state of *trans*-octatetraene and 10 vibrational sublevels of *cis*-octatetraene corresponding to totally symmetrical vibrations with frequencies 0 cm$^{-1}$ < $\omega$ < 2000 cm$^{-1}$. This leads to another noticeable distinction in isomer spectra; specifically, lines with frequencies greater than the frequency of the pure electron transition are present in the spectrum of *cis*-octatetraene, which is a distinctive feature of this component in the mixture spectrum along with the time dependence.

Similar results were obtained for other molecules of the polyene series. For example, the difference between the intensities of lines of stationary spectra of various isomeric forms of butadiene does not exceed 20% and time dependences of their spectra differ by a factor of 3, which ensures their reliable interpretation in the range of relative concentrations from 1 to 110%. As the molecular size increases, spectral effects associated with isomerization become smaller as a rule. This is caused by a decrease in the relative effect of local variations of the molecular structure (modes of vibrations and electron density) on the probabilities and the transition energies. This is manifested particularly well in diphenyl-substituted polyenes.

Thus, stationary spectra of rotational isomers of diphenyloctatetraene (DO) differ only in details, namely, in the presence of low-intensity lines with relative intensities comparable with the accuracy of their calculation ($\leq 10\%$, Fig. 1b). Although the type of isomerization here is the same as in the case of octatet-raene considered above, there are no active lines in the spectra whose intensities differ significantly for the two given isomers. Simulation of time-resolved spectra showed that calculated characteristics of the first electron-excited states and transitions from them to the ground states are relatively close for *cis*- and *trans*- diphenyloctatetraene (the lifetimes are 90 and 60 ps, the oscillator strengths are 1.5 and 2.2, and the energies are 29 600 and 30 500 cm$^{-1}$, respectively) and the spectrum of a mixture of isomers excited to these states (Fig. 3) consists of two strongly overlapping components with similar temporal dependences. Even in this situation, lines can be assigned to individual components of the mixture and the presence of each isomeric form in the mixture can be determined although with a lower accuracy, i.e., in a smaller range of possible relative concentrations ($C_{cis\text{-DO}}/C_{trans\text{-DO}} = 0.2$–$20$).

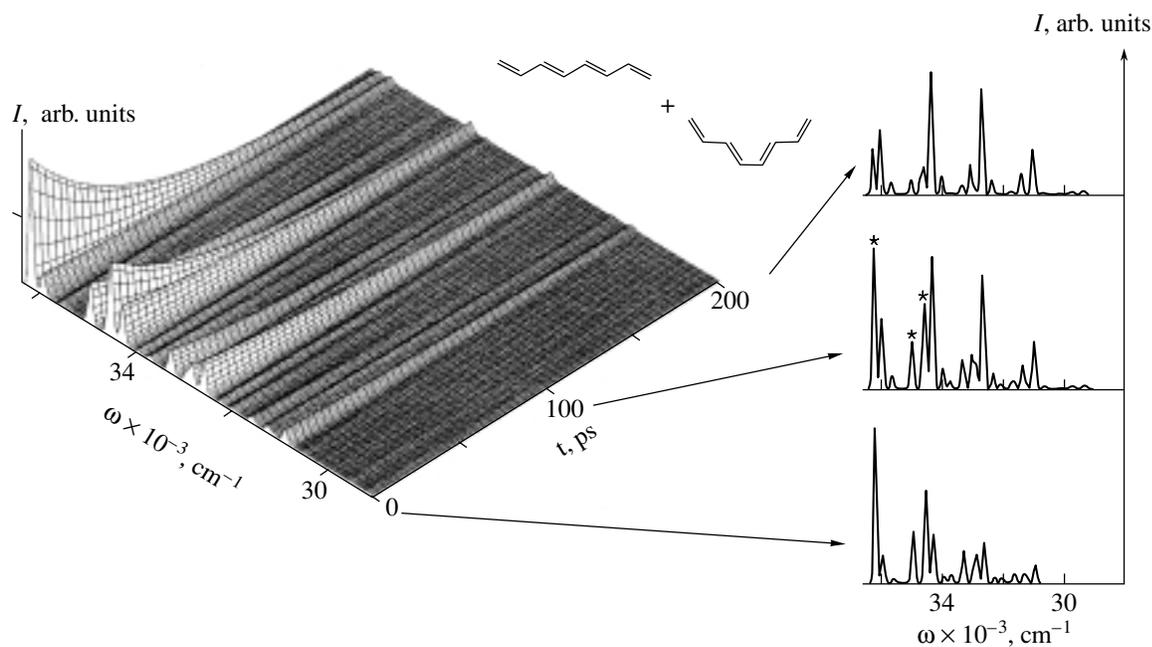

**Fig. 2.** Calculated time-resolved fluorescence spectra of a mixture of rotational isomers of octatetraene.

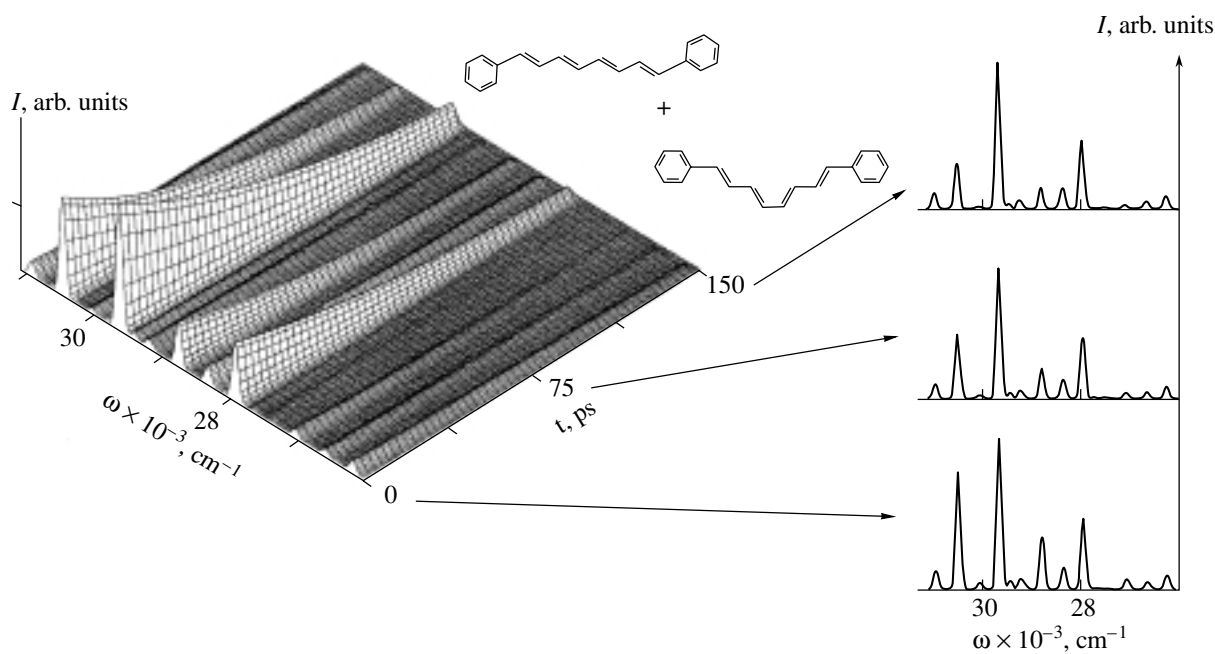

**Fig. 3.** Calculated time-resolved fluorescence spectra of a mixture of rotational isomers of diphenyloctatetraene.

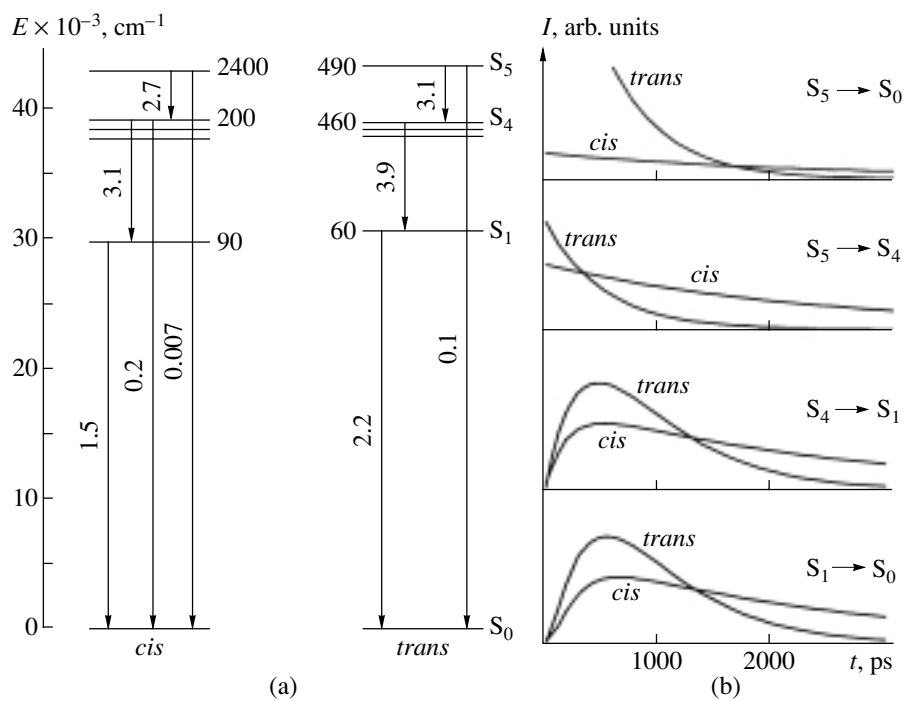

**Fig. 4.** (a) Energy level diagram and (b) time dependences of fluorescence intensities of models of isomers of a diphenyloctatetraene molecule. Transition oscillator strengths and level lifetimes in picoseconds are presented in the diagram.

Results of computer simulation showed that the efficiency of analysis can be significantly improved by a somewhat different arrangement of the experiment. The calculations show that the lifetimes of lower excited states of diphenyloctatetraene differ relatively slightly, by a factor of ~1.5, whereas highly excited states are more sensitive to variations in the electron structure associated with isomerization. This is clearly seen (Fig. 4a) from differences between oscillator strengths of the transitions and lifetimes of the levels (2–5 times for the states $S_4$ and $S_5$) whereas their energies coincide to an accuracy on the order of the energy of a vibrational quantum. Therefore, isomers will emit in the same frequency regions in transitions from highly-excited electron states (the vibrational structures of the spectra will overlap) but time dependences of the intensities of their lines will be radically different. This provides conditions for spectral analysis and determination of the composition of mixtures of these isomers from time-resolved spectra recorded in a wide spectral range (0 cm$_{-1}$ < $\omega$ < 50 000 cm$^{-1}$). Time dependences of the intensities of 0–0 components for the most intense electron transitions that can be recorded in the case of excitation of the fifth electron state $S_5$ of rotational isomers of diphenyloctatetraene are shown in Fig. 4b. It is clearly seen that time dependences in this excitation are heavily different not only for the spectra corresponding to transitions from highly excited states ($S_5 \longrightarrow S_0$, $S_5 \longrightarrow S_4$, and $S_4 \longrightarrow S_1$) but also for the spectrum of $S_1 \longrightarrow S_0$. These distinctions are approximately three times larger than the case of excitation to the state $S_1$, while the range of possible relative concentrations $C_{cis\text{-DO}}/C_{trans\text{-DO}}$ increases up to 2.5–250, i.e., approximately tenfold.

Similar data were obtained for other molecules considered. This feature is due to a significantly higher sensitivity of high electron-excited states to isomer restructuring as compared to lower states and demonstrates a higher analytical efficiency of methods of time-resolved spectroscopy in the case of excitation to higher electron states.

The presence of substituents of various types in a molecule is represented in the spectra in varying degrees. For instance, the substitution effect can be detected reasonably well in the comparison of stationary spectra of diphenylbutadiene and butadiene [5] (lines 1200 and 1600 cm$^{-1}$ in the spectra of substituted molecules are approximately 10–40% less intense), diphenyloctatetraene (Fig. 1b), and octatetraene (Fig. 1a) (the difference between the intensities of the main lines is ~20% for *cis*- and *trans*-forms). However, these differences for methyl-substituted octatetraene (decatetraene) in conventional spectra are 5–10%, which is comparable with the accuracy of calculation of the intensities, and do not allow reliable identification. Calculated parameters of time dependences of octatetraene and decatetraene spectra for transitions from the lower electron-excited state are also close to each other (50 and 55 ps). Nevertheless, estimates made in [18] show that a confident quantitative analysis can be made in this case as well but for a longer time of spectrum observation, on the order of 500 ps. Note that the presence of a difference between the time dependences can be detected in a much shorter time and that this additional information allows quantitative identification of the presence of mixture components.

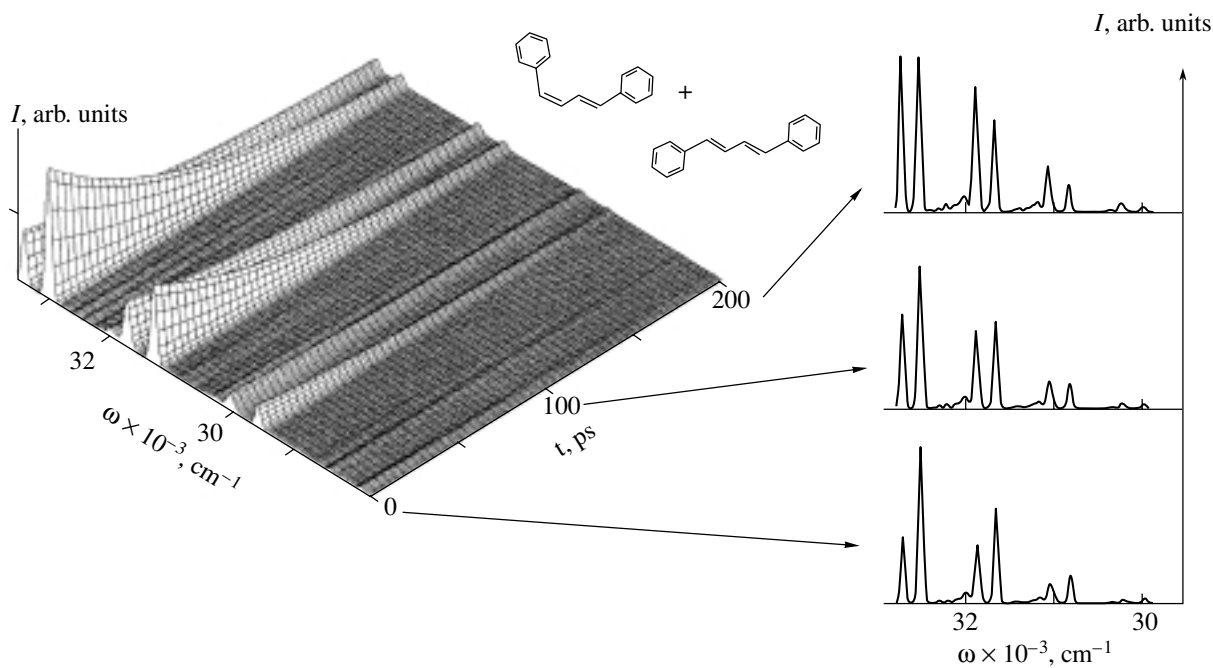

**Fig. 5.** Calculated time-resolved fluorescence spectra of a mixture of rotational isomers of phenyl-substituted butadiene.

The use of time-resolved spectra can be rather efficient also in the case when it is necessary to identify molecules with different locations of substituents and it is difficult to identify them from ordinary spectra. For example, calculated stationary fluorescence spectra of *cis*- and *trans*-substituted butadiene (Fig. 1c) are difficult to distinguish (the difference is no larger than 10%) and the difference between the energies of pure electron transitions is too small to be determined (33400 and 33000 cm$^{-1}$). The similarity of the spectra is caused first of all by the fact that these molecules differ only in the location of substituents (phenyl rings) whereas their polyene chains are the same. As for the spectrum structure (line frequencies and intensities), it is determined by the difference between parameters of molecular models (lengths of bonds and force constants) of the lower excited and the ground states. Calculations show that variations in the electron density resulting from the transition to the first excited state of diphenylpolyenes (and diphenylbutadiene in particular) occur mainly in the polyene part of molecules and, correspondingly, distinctions for bonds in the polyene parts are 2–3 times greater than in phenyl rings. Therefore, ordinary spectra of phenyl-substituted butadiene with identical polyene chains are practically independent of the locations of substituents.

Simulation of the three-dimensional spectra of a mixture of substituted butadiene (Fig. 5) showed that molecules with different locations of phenyl rings can be identified from time dependences of fluorescence line intensities, which are different. Lifetimes of the lower excited states are 120 and 80 ps for substituents in *cis*- and *trans*-locations. In spite of the fact that spectra of components strongly overlap, lines with different decay rates can be identified using sections of the profile of the mixture spectrum. As in the case of isomers of diphenyloctatetraene considered above, differences in time dependences of components of the mixture spectrum will be heavier if molecules are excited to high-lying electron states.

## CONCLUSIONS

Three-dimensional spectra of isomers and substituted molecules in this work are calculated with the use of molecular models with parameters obtained previously from the comparison of their stationary spectra with experiment. We also used the approximation consisting in neglect of nonradiative transitions (vibrational relaxation, internal conversion and intersystem crossing), transitions between isomers, and electron–vibration interaction. All these factors, which often play an important role in femtosecond spectroscopy, can noticeably affect the form of the spectrum both qualitatively (multiexponential time dependences) and quantitatively. However, by virtue of the fact that the isomers considered and the substituted molecules have similar structures, we can reasonably expect a similar influence of the effects neglected in this analysis on the quantities responsible for the shape of time-resolved spectra (transition probabilities and energies). Thus, for instance, the lifetime of the lower electron state of *trans*-diphenylbutadiene is estimated from experimental data [20] as 345 ps (80 ps in the calculation). The disagreement can be associated, for instance, with vibronic interaction (neglected in the Franck–Condon model used) between a given electron state of the $B_u$ symmetry with another closely-lying state of $A_g$ symmetry, from which the transition to the ground state is forbidden. The result of this interaction is governed by the relative location of energy levels and oscillator strengths of the transitions, which differ little for the rotational isomers and substituted molecules compared. Therefore, the quantitative estimates of temporal characteristics obtained can experience significant changes as we pass to more accurate models, but the pattern obtained does not radically change and includes the possibility for confident interpretation as revealed in numerical experiments and estimates of relative concentrations of mixture components.

The results obtained demonstrate the efficiency of methods of time-resolved spectroscopy in spectral analysis of chemical compounds with similar optical properties when traditional methods (stationary spectra) are inefficient. For improving the reliability and accuracy of the results obtained in computer simulation the molecular models and approximations used should be refined, including the solution of inverse problems. These studies are presently being carried out. In particular, methods for calculating nonradiative transition probabilities, ways and probabilities of interisomer conversions and chemical reactions [21–25] are being studied. In combination with modern potentials of spectral experiments with femtosecond time resolution (including the case of interisomer restructuring [2]) this makes active application of software for simulating the three-dimensional spectra of complex molecules developed here highly promising in qualitative and quantitative analysis.

The technique of femtosecond spectroscopy is rather complicated in itself and the production of dynamic spectra is far from elementary, particularly if it is aimed at solving particular analytical problems for complex molecules. Therefore, the possibility for reasonable planning of such experiments and their initial execution under optimum conditions adequate to the particular problem solved are extremely important. This can be made only by using a numerical experiment. The study carried out showed that modern achievements of theoretical optical spectroscopy can be safely used as the basis for performing large-scale computer simulation with the aim of optimizing the performance of natural experiments.

*Translated by A. Mozharovskiĭ*